\documentclass[aps,prl,superscriptaddress,amsmath,twocolumn,nofootinbib,showpacs]{revtex4-1}

\usepackage{amssymb}
\usepackage{graphicx}
\usepackage[hang,nooneline]{subfigure}
\usepackage{dcolumn}
\usepackage{bm}
\usepackage[pdftex]{color}
\usepackage[sort&compress]{natbib}
\usepackage[colorlinks=true,linkcolor=blue,filecolor=blue,urlcolor=blue,citecolor=blue,pdftex=true,
plainpages=false]{hyperref}

\newcommand{\BR}{\ensuremath{\text{BR}}}   

\definecolor{blue}{RGB}{0,51,133}
\definecolor{green}{RGB}{81,134,36}
\definecolor{cyan}{rgb}{0,0.8,0.8}
\definecolor{purple}{rgb}{0.7,0,0.7}
\definecolor{orange}{RGB}{238,80,25}
\definecolor{brown}{RGB}{133,102,0}

\newcommand{\snip}[1]{}

\begin{document}
\author{Jon A. Bailey}
\affiliation{Department of Physics and Astronomy, Seoul National University, Seoul, South Korea}

\author{A.~Bazavov}
\affiliation{Department of Physics and Astronomy, University of Iowa, Iowa City, IA, USA}

\author{C.~Bernard}
\affiliation{Department of Physics, Washington University, St.~Louis, MO, USA}

\author{C.M.~Bouchard}
\affiliation{Department of Physics, The Ohio State University, Columbus, OH, USA}

\author{C.~DeTar}
\affiliation{Department of Physics and Astronomy, University of Utah, Salt Lake City, UT, USA}

\author{Daping Du}
\email[]{dadu@syr.edu}
\affiliation{Department of Physics, Syracuse University, Syracuse, NY, USA}

\author{A.X.~El-Khadra}
\affiliation{Department of Physics, University of Illinois, Urbana, IL, USA}

\author{E.D.~Freeland}
\affiliation{Liberal Arts Department, School of the Art Institute of Chicago, Chicago, IL, USA}

\author{E.~G\'amiz}
\affiliation{CAFPE and Departamento de Fisica Te\'orica y del Cosmos, Universidad de Granada, Granada, Spain}

\author{Steven~Gottlieb}
\affiliation{Department of Physics, Indiana University, Bloomington, IN, USA}

\author{U.M.~Heller}
\affiliation{American Physical Society, Ridge, NY, USA}

\author{A.S.~Kronfeld}
\affiliation{Fermi National Accelerator Laboratory, Batavia, IL, USA}
\affiliation{Institute for Advanced Study, Technische Universit\"at M\"unchen, Garching, Germany}

\author{J.~Laiho}
\affiliation{Department of Physics, Syracuse University, Syracuse, NY, USA}

\author{L.~Levkova}
\affiliation{Department of Physics and Astronomy, University of Utah, Salt Lake City, UT, USA}

\author{Yuzhi Liu}
\affiliation{Department of Physics, University of Colorado, Boulder, CO, USA}

\author{E.~Lunghi}
\email[]{elunghi@indiana.edu}
\affiliation{Department of Physics, Indiana University, Bloomington, IN, USA}

\author{P.B.~Mackenzie}
\affiliation{Fermi National Accelerator Laboratory, Batavia, IL, USA}

\author{Y. Meurice}
\affiliation{Department of Physics and Astronomy, University of Iowa, Iowa City, IA, USA}

\author{E.~Neil}
\affiliation{Department of Physics, University of Colorado, Boulder, CO, USA}
\affiliation{RIKEN-BNL Research Center, Brookhaven National Laboratory, Upton, NY, USA}

\author{Si-Wei~Qiu}
\affiliation{Department of Physics and Astronomy, University of Utah, Salt Lake City, UT, USA}

\author{J.N.~Simone}
\affiliation{Fermi National Accelerator Laboratory, Batavia, IL, USA}

\author{R.~Sugar}
\affiliation{Department of Physics, University of California, Santa Barbara, CA, USA}

\author{D.~Toussaint}
\affiliation{Physics Department, University of Arizona, Tucson, AZ, USA}

\author{R.S.~\surname{Van de Water}}
\email[]{ruthv@fnal.gov}
\affiliation{Fermi National Accelerator Laboratory, Batavia, IL, USA}

\author{Ran~Zhou}
\affiliation{Fermi National Accelerator Laboratory, Batavia, IL, USA}

\collaboration{Fermilab Lattice and MILC Collaborations}
\noaffiliation

\preprint{FERMILAB-PUB-15-288-T}

\title{\boldmath $B\to\pi\ell\ell$ form factors for new-physics searches from lattice QCD}

\date{\today} 

\begin{abstract}
The rare decay $B\to\pi\ell^+\ell^-$ arises from $b\to d$ flavor-changing neutral currents and could be sensitive to physics beyond the Standard Model.
Here, we present the first \emph{ab-initio} QCD calculation of the $B\to\pi$ tensor form factor $f_T$.
Together with the vector and scalar form factors $f_+$ and $f_0$ from our companion work~%
[J.~A.\ Bailey \emph{et al.}, Phys.\ Rev.\ D  {\bf 92}, 014024 (2015)],
these parameterize the hadronic contribution to $B\to\pi$ semileptonic decays in any extension of the
Standard Model.
We obtain the total branching ratio $\BR(B^+\to\pi^+\mu^+\mu^-)=20.4(2.1)\times10^{-9}$ in the Standard
Model, which is the most precise theoretical determination to date, and agrees with the recent measurement
from the LHCb experiment~%
[R.~Aaij \emph{et al.}, JHEP {\bf 1212}, 125 (2012)].
Note added: after this paper was submitted for publication, LHCb announced a new measurement of the differential decay rate for this process [T. Tekampe, talk at DPF 2015], which we now compare to the shape and normalization of the Standard-Model prediction.
\end{abstract}

\pacs{13.20.He, 
      12.38.Gc, 
      12.15.Mm\vspace{-3pt}}  

\maketitle



\emph{Motivation} --- %
Hadron decays that proceed through flavor-changing neutral currents may be sensitive to new physics, because
their leading Standard-Model contributions are loop suppressed.
Here we study the semileptonic decay $B\to\pi\ell^+\ell^-$, which proceeds through a $b\to d$ transition.
Hadronic effects in this decay are parametrized by three form factors.
In this Letter, we present the first \emph{ab-initio} QCD calculation of the tensor form factor $f_T$, based
on lattice-QCD work that also yielded the vector and scalar form factors, $f_+$ and
$f_0$~\cite{Lattice:2015tia}.
Lattice QCD has several advantages over other approaches to the form
factors~\cite{Ball:2004ye,Wang:2007sp,Duplancic:2008ix,Wu:2009kq,Faustov:2014zva,Ali:2013zfa,Li:2014uha,Hambrock:2015wka},
particularly in providing a path to controlled uncertainties that can be systematically
reduced~\cite{Kronfeld:2012uk}.

The LHCb experiment recently made the first observation of $B^+\to\pi^+\mu^+\mu^-$~\cite{LHCb:2012de}, while
the $B$-factories have set limits on the $e^+e^-$ and $\tau^+\tau^-$
channels~\cite{Wei:2008nv,Lees:2013lvs,Lutz:2013ftz}.
Below we present the first calculations of $B\to\pi\ell^+\ell^-$ ($\ell=e,\mu,\tau$) observables in the
Standard Model using form factors with fully controlled uncertainties.

The form factors $f_+$, $f_0$, and $f_T$ suffice to parameterize $B\to\pi$ decays in all extensions of the
Standard Model.
New physics from heavy particles---such as those appearing in models with 
supersymmetry~\cite{Bobeth:2001sq,Demir:2002cj,Choudhury:2002fk,Wang:2007sp},
a fourth generation~\cite{Hou:2013btm},
or extended~\cite{Aliev:1998sk,Iltan:1998ra,Bobeth:2001sq,Erkol:2002nw,Erkol:2004me,Song:2008zzc} or
composite~\cite{Gripaios:2014tna} Higgs sectors---alter Wilson coefficients in the effective Hamiltonian
pertaining to particle physics below the electroweak
scale~\cite{Grinstein:1988me,Buras:1993xp,Huber:2005ig,Altmannshofer:2008dz}.
Whatever these unknown particles may be, the hadronic physics remains the same.


\emph{Lattice-QCD calculation} --- %
Our work on $f_T(q^2)$ was carried out in parallel with $f_+(q^2)$ and $f_0(q^2)$.
Our aim in Ref.~\cite{Lattice:2015tia} was a precise determination of the Cabibbo-Kobayashi-Maskawa (CKM)
element $|V_{ub}|$, and every step of the analysis was subjected to many tests.
Further, two of the authors applied a multiplicative offset to the numerical data at an early stage.
This ``blinding'' factor was disclosed to the others only after finalizing the error analysis.
Full details of the simulation parameters, analysis, and cross-checks are given in
Ref.~\cite{Lattice:2015tia}.

Our calculation uses ensembles of lattice gauge-field configurations 
\cite{asqtad:en06a,*asqtad:en06b,*asqtad:en05a,*asqtad:en05b,*asqtad:en04a,
*asqtad:en15a,*asqtad:en15b,*asqtad:en14a,*asqtad:en13a,*asqtad:en13b,*asqtad:en12a,
*asqtad:en23a,*asqtad:en23b,*asqtad:en20a,*asqtad:en20b,*asqtad:en19a,*asqtad:en18a,*asqtad:en18b,
*asqtad:en24a} 
from the MILC Collaboration~\cite{Bernard:2001av,Aubin:2004wf,Bazavov:2009bb}, which are generated with a
realistic sea of up, down, and strange quarks.
In practice, the up and down sea quarks have the same mass, and the strange-quark mass is tuned close to its
physical value.
The statistics are high, with 600--2200 gauge-field configurations per ensemble.
The physical volume is large enough that we can repeat the calculation in different parts of the lattice,
thereby quadrupling the statistics.
We use four lattice spacings ranging from 0.12~fm to 0.045~fm to control the extrapolation to zero lattice
spacing.

The tensor form factor is defined via the matrix element of the $b \to d$ tensor current
$i\bar{d}\sigma^{\mu\nu}b$:
\begin{equation}
    \langle\pi(p_\pi)|i\bar{d}\sigma^{\mu\nu}b|B(p_B)\rangle =
        2\frac{p_B^\mu p_\pi^\nu - p_B^\nu p_\pi^\mu}{M_B+M_\pi}\,f_T(q^2),
    \label{eq:fT}
\end{equation}
where $p_B$ and $p_\pi$ are the particles' momenta and $q=p_B-p_\pi$ is the momentum carried off by the
leptons.
The Lorentz invariant $q^2$ is related to the pion energy in the $B$-meson rest frame via
$E_\pi=(M_B^2+M_\pi^2-q^2)/2M_B$.
In the finite volume that can be simulated on a computer, $E_\pi$ takes discrete values, dictated by the
spatial momenta~$\bm{p}_\pi$ compatible with periodic boundary conditions.
Because statistical and discretization errors increase with pion momentum, we restrict
$|\bm{p}_\pi|\leq|2\pi(1,1,1)/L|$.
The resulting simulation range of $E_\pi\lesssim1$~GeV is significantly smaller than the kinematically
allowed range of $E_\pi\leq2.5$~GeV.
Extending this discrete set of calculations into the full $q^2$ dependence is the central challenge of this
work, and is met in two steps.

The two light quarks (up and down) have a mass larger than it should be, but the range simulated is wide and
the smallest pion mass is 175~MeV, close to Nature's 140~MeV.
Therefore, we can apply an effective field theory of pions---chiral perturbation theory---to extrapolate the
simulation data to the physical point.
We use a form of chiral perturbation theory adapted to lattice QCD, with additional terms describing the
lattice-spacing dependence~\cite{Aubin:2007mc,RanBtoK} and with modifications needed for energetic
final-state pions~\cite{Bijnens:2010ws}.
As discussed in Ref.~\cite{Lattice:2015tia}, we try several fit variations.
For example, we replace the loop integrals with momentum sums appropriate to the finite volume, finding
negligible changes in the results.
Our final fit includes next-to-next-to-next-to-leading order analytic terms and terms to model the
discretization errors of the heavy quark.
The latter come from an effective field theory for heavy $b$
quarks~\cite{Kronfeld:2000ck,Harada:2001fi,Oktay:2008ex}.

\begin{figure}[b]
    \includegraphics[width=0.98\linewidth]{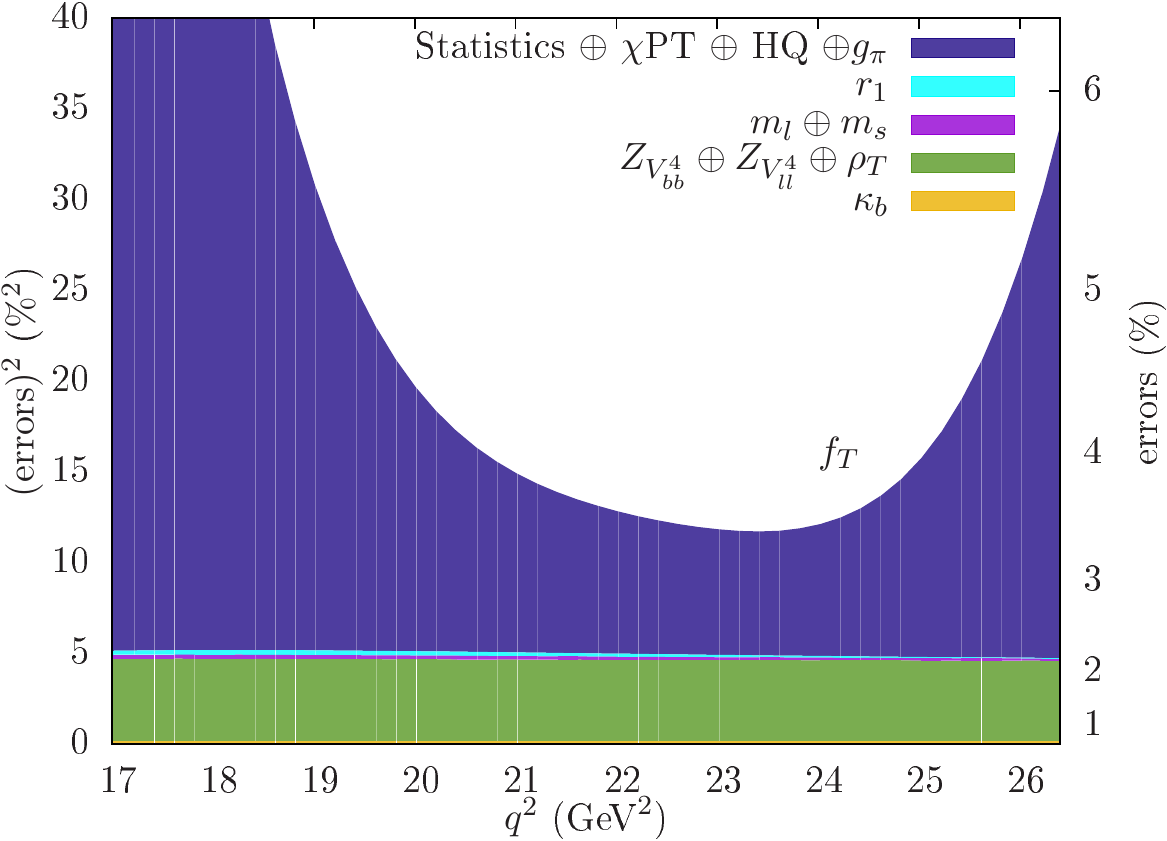}
    \caption{(color online) Error budget for $f_T$ as a function of~$q^2$ for the range of simulated
        lattice momenta.
        The filled bands show the relative size of each error contribution to the total.
        The quadrature sum is shown on the left $y$~axis and the error itself, in per cent, on the right.}
    \label{fig:fT_Error}
\end{figure}

Figure~\ref{fig:fT_Error} shows the $q^2$ dependence of the errors after the chiral-continuum extrapolation
just described.
Table~\ref{tab:fT_Error} gives a numerical error budget for $f_T(q^2=20~\text{GeV}^2)$.
\begin{table}
    \caption{Error budget in per cent for $f_T(q^2=20~\text{GeV}^2)$.
        The first error incorporates statistical errors from the simulation and systematics
        associated with the chiral-continuum fit.
        The last column emphasizes how the error varies with~$q^2$.}
    \label{tab:fT_Error}
    \begin{tabular}{l@{\quad}c@{\quad}c}
        \hline\hline
        Source of error                                             & $\delta f_T$ & $q^2$ dependence \\
        \hline
        Statistics $\oplus$ $\chi$PT $\oplus$ HQ $\oplus$ $g_{\pi}$ &      3.8     & important  \\
        Scale $r_1$                                                 &      0.5     & negligible \\
        Nonperturbative matching $Z_{V_{bb}^4}, Z_{V_{ll}^4}$       &      0.7     & negligible \\
        Perturbative matching $\rho_{T}$                            &      2.0     & none  \\
        Heavy-quark mass tuning $\kappa_b$                          &      0.4     & none  \\
        Light-quark mass tuning $m_l, m_s$                          &      0.5     & negligible \\
        \hline
        Total (\emph{Quadrature sum of above})                      &      4.4     & important  \\
        \hline\hline
    \end{tabular}
\end{table}
The largest uncertainty comes from the statistical errors, as increased during the chiral-continuum
extrapolation.
This error is under good control for $q^2$ corresponding to the spatial momenta that we simulate, but grows
large elsewhere.

The subdominant errors are as follows.
To convert from lattice units to physical units, we introduce a physical distance $r_1$, which is defined
via the force between static quarks~\cite{Bernard:2000gd,Sommer:1993ce}.
We use it to form physical, dimensionless quantities, which are the input data for the chiral-continuum fit.
At the end, we set $r_1=0.3117\pm0.0022$~fm \cite{Bazavov:2011aa} based on a related lattice-QCD calculation
of $r_1f_\pi$~\cite{Bazavov:2009fk} and the pion decay constant $f_\pi=130.41$~MeV~\cite{Agashe:2014kda}.
To propagate the parametric uncertainty in $r_1$ to $f_T$, we repeat the fit shifting $r_1$ by
$\pm1\sigma_{r_1}$, leading to the second line in Table~\ref{tab:fT_Error}.

In lattice gauge theory, the tensor current does not have the normalization used in QCD phenomenology.
We obtain most of the normalization nonperturbatively~\cite{ElKhadra:2001rv} from $b\to b$ and $d\to d$
transitions with the vector current, with statistical errors below~1\%.
Another matching factor $\rho_T$ remains, but, by design and in practice, it is close to unity.
We calculate $\rho_T$ at the renormalization scale $\mu=m_{b,\text{pole}}$ through first order in the QCD
coupling $\alpha_s$.
We estimate the resulting error of order~$\alpha_s^2$ after removing a logarithmic dependence on the
matching scale~$\mu$, which is present in continuum QCD too.
We then examine how the one-loop coefficient depends on heavy-quark mass, identifying the largest value,
$\rho^{[1]}_{T,\,\text{max}}$.
Finally, we estimate the error in $\rho_T$ to be $2\alpha_s^2|\rho^{[1]}_{T,\,\text{max}}|$, evaluating
$\alpha_s$ on the second-finest lattice with $a\approx0.06$~fm.
This yields the 2\% perturbative-matching uncertainty in Table~\ref{tab:fT_Error}.

The last two uncertainties arise as follows.
When generating data, we choose the simulation quark masses based on short runs and previous experience.
The full analysis yields better estimates.
To correct the simulation $b$-quark mass \emph{a posteriori}, we recompute $f_T$ on one ensemble with two
additional values of the bare $b$-quark mass.
Using the slope from all three mass values, we interpolate the data for $f_T$ slightly from the production
$b$-quark mass to the physical value.
This leaves an error due to the uncertainty in the size of the $b$-quark mass correction.
The details for $f_T$ are nearly identical to those for $f_+$~\cite{Lattice:2015tia}, leading to the same
estimate, 0.4\%, for this error.
The light-quark mass dependence is embedded in the chiral-continuum extrapolation, described above.
The parametric uncertainty from the input light-quark mass~\cite{Bazavov:2009bb} is propagated to $f_T$ by
repeating the fit with $\pm1\sigma_{m_q}$ shifts to these parameters, and is given in the penultimate line
of Table~\ref{tab:fT_Error}.

The final line in Table~\ref{tab:fT_Error} and the upper edge of the stack in Fig.~\ref{fig:fT_Error}
represent the quadrature sum of the systematic uncertainties with the chiral-continuum fit error.


\emph{Extension to all $q^2$} --- %
To extend $f_T$ in the chiral-continuum limit from the range of simulated lattice momenta to the full
kinematic range, $0<q^2\le(M_B-M_\pi)^2$, with controlled errors, we use a method based on the analytic
structure of the form factor.

In the complex $q^2$ plane, $f_T(q^2)$ has a cut for timelike $q^2\ge t_+\equiv(M_B+M_\pi)^2$ and a pole at
$q^2=M_{B^*}^2$ but is analytic elsewhere.
The variable
\begin{equation}
    z(q^2,t_{0}) = \frac{\sqrt{t_{+}-q^2}-\sqrt{t_{+}-t_{0}}}{\sqrt{t_{+}-q^2 }+\sqrt{t_{+}-t_{0}}}
    \label{eq:Z(t)}
\end{equation}
maps the whole $q^2$ plane into the unit disk, with the cut mapped to the boundary and the semileptonic
region mapped to an interval on the real axis.
Unitarity bounds then guarantee that an expansion of $f_T$ in $z$ (with the $B^*$ pole removed) converges
for $|z|<1$~\cite{Bourrely:1980gp,Boyd:1994tt,Lellouch:1995yv,Boyd:1997qw}.
Following Bourrely, Caprini, and Lellouch (BCL)~\cite{Bourrely:2008za}, we factor out the $B^*$ pole and
expand in $z$:
\begin{equation}
    (1-q^2/M_{B^*}^2) f_T(z) = \sum_{n=0}^{N_z-1}b^T_{n}
        \left[ z^{n} - (-1)^{n-N_z}\frac{n}{N_z}z^{N_z} \right],
    \label{eq:fT(z)}
\end{equation}
choosing $t_0=(M_B+M_\pi)(\sqrt{M_B}-\sqrt{M_\pi})^2$ to minimize $|z|$ in the semileptonic region.
Although Eq.~(\ref{eq:fT(z)}) was derived for the vector form factor $f_+$, we use it for the tensor form
factor $f_T$ because the two form factors are proportional to each other at leading order in the $1/m_b$
expansion.

\begin{figure}[b]
    \includegraphics[width=0.98\linewidth]{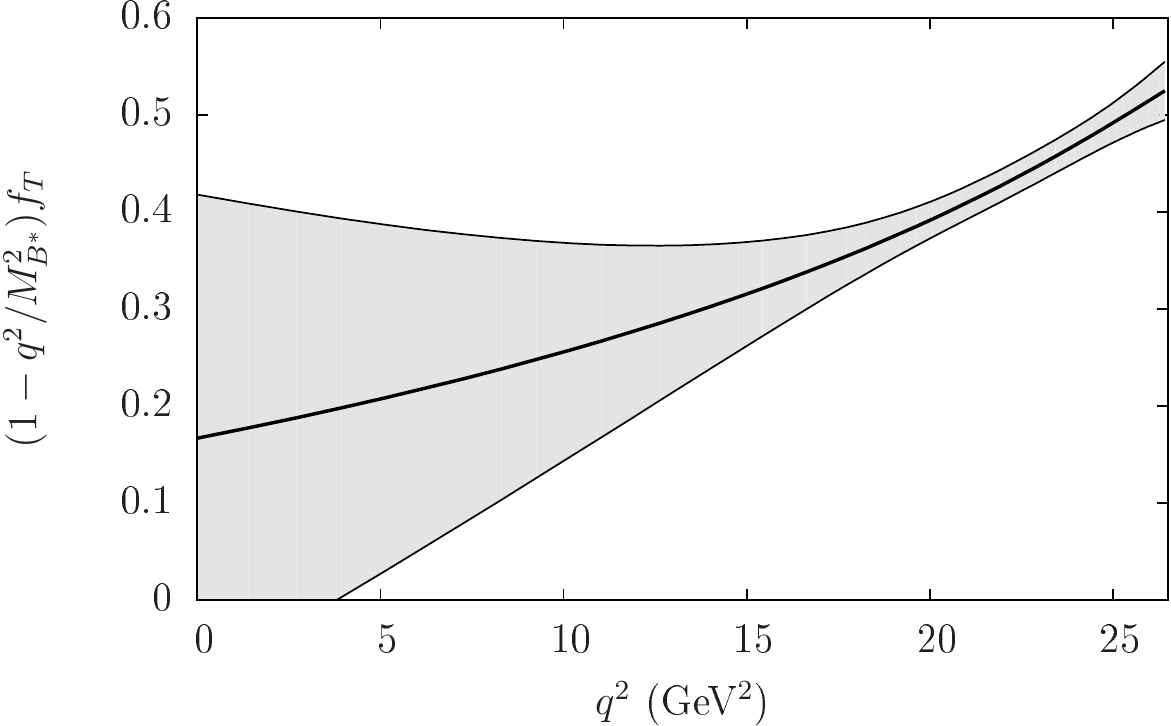}
         \caption{\emph{Ab-initio} result for $f_T(q^2)$ from lattice QCD.}
    \label{fig:fT_zFit}
\end{figure}

We determine the $b_n^T$ with a functional method connecting the independent functions of the
chiral-continuum fit with the first several powers of~$z$~\cite{Lattice:2015tia}.
Our preferred fit uses $N_z=4$; adding higher-order terms in $z$ does not significantly change the central
value.
Table~\ref{tab:fT_correlation} presents our final result for $f_T$ as coefficients of the $N_z=4$ BCL
$z$~fit and the correlation matrix between them, where the errors include statistical and all systematic
uncertainties.
\begin{table}
    \caption{Best-fit values $b_n$ with total errors and correlation matrix $\rho_{nm}$ of the $N_z=4$ BCL
        $z$~expansion of $f_T$.
        The lower two panels show correlations with the $f_+$ and $f_0$ coefficients in Table~XIV of
        Ref.~\cite{Lattice:2015tia} obtained from \emph{ab-initio} QCD.}
    \label{tab:fT_correlation}
    \begin{tabular}{lcccc}
        \hline \hline
        Fit:        & 0.393(17) & $-0.65(23)$ & $-0.6(1.5)$ & 0.1(2.8) \\
        \hline
        $\rho$      &  $b_0^T$  &   $b_1^T$   &   $b_2^T$   &  $b_3^T$ \\
        \hline
        $b_{0}^{T}$ &   1.000   &    0.400    &    0.204    &   0.166  \\
        $b_{1}^{T}$ &           &    1.000    &    0.862    &   0.806  \\
        $b_{2}^{T}$ &           &             &    1.000    &   0.989  \\
        $b_{3}^{T}$ &           &             &             &   1.000  \\
        \hline
        $b_{0}^{+}$ &   0.638   &    0.321    &    0.123    &    0.084  \\
        $b_{1}^{+}$ &   0.321   &    0.397    &    0.162    &    0.109  \\
        $b_{2}^{+}$ &   0.114   &    0.202    &    0.198    &    0.179  \\
        $b_{3}^{+}$ &   0.070   &    0.152    &    0.192    &    0.180  \\
        \hline
        $b_{0}^{0}$ &   0.331   &    0.136    &    0.089    &    0.073  \\
        $b_{1}^{0}$ &   0.203   &    0.313    &    0.198    &    0.162  \\
        $b_{2}^{0}$ &   0.204   &    0.268    &    0.186    &    0.155  \\
        $b_{3}^{0}$ &   0.151   &    0.203    &    0.169    &    0.149  \\
        \hline \hline
    \end{tabular}
\end{table}
This information can be used to reconstruct $f_T(q^2)$ over the full kinematic range.
Table~\ref{tab:fT_correlation} also provides the (mostly statistical) correlations between $f_T$, $f_+$,
and~$f_0$.
Figure~\ref{fig:fT_zFit} shows the extrapolation of $f_T$ to $q^2=0$.
Table~\ref{tab:fT_correlation} and Fig.~\ref{fig:fT_zFit} represent the first main result of this Letter.


\emph{Implications} --- %
The largest contribution in the Standard Model to the amplitude for $B\to\pi\ell^+\ell^-$ is proportional to
the vector form factor.
Assuming that new physics does not contribute significantly to the tree-level decay $B\to\pi\ell\nu$, one
can use experimental measurements of this process to constrain the shape of $f_+(q^2)$, especially at
low~$q^2$.
In Ref.~\cite{Lattice:2015tia}, we obtain the CKM element $|V_{ub}|$ from a combined $z$~fit to our
lattice-QCD results for $f_+$ and $f_0$ and measurements of $\tau_Bd\Gamma(B\to\pi\ell\nu)/dq^2$ from
BaBar~\cite{delAmoSanchez:2010af,Lees:2012vv} and Belle~\cite{Ha:2010rf,Sibidanov:2013rkk}.
This joint fit also yields the most precise current determinations of $f_+$ and $f_0$.
To enable them to be combined with the results for $f_T$ from Table~\ref{tab:fT_correlation},
Table~\ref{tab:best_fT_correlation} provides the correlations between the $z$-expansion coefficients for all
three form factors.
\begin{table}
    \caption{Correlations between BCL coefficients for $f_T$ with those
        for $f_+$ and $f_0$ from Table~ XIX of Ref.~\cite{Lattice:2015tia}, which include
        experimental shape information from $B\to\pi\ell\nu$ decay.}
    \label{tab:best_fT_correlation}
    \begin{tabular}{lrrrr}
        \hline \hline
        $\rho$      & $b_0^T$  & $b_1^T$  & $b_2^T$  & $b_3^T$  \\
        \hline
	$b_{0}^{+}$ & 0.514 & 0.140 & 0.078 & 0.065 \\		 
	$b_{1}^{+}$ & 0.111 & 0.221 & $-$0.010 & $-$0.049 \\	
	$b_{2}^{+}$ & $-$0.271 & $-$0.232 & $-$0.012 & 0.029 \\	
	$b_{3}^{+}$ & $-$0.204 & $-$0.215 & $-$0.013 & 0.023 \\ 
        \hline
	$b_{0}^{0}$ & 0.243 & $-$0.015 & $-$0.025 & $-$0.024 \\		 
	$b_{1}^{0}$ & 0.005 & 0.134 & 0.070 & 0.057 \\	
	$b_{2}^{0}$ & $-$0.002 & $-$0.034 & $-$0.032 & $-$0.030 \\	
	$b_{3}^{0}$ & $-$0.044 & $-$0.061 & 0.005 & 0.017 \\ 
        \hline \hline
    \end{tabular}    
\end{table}
The correlations are small, because $f_+$ contains independent experimental information.

\begin{figure}[t]
    \includegraphics[width=0.98\linewidth]{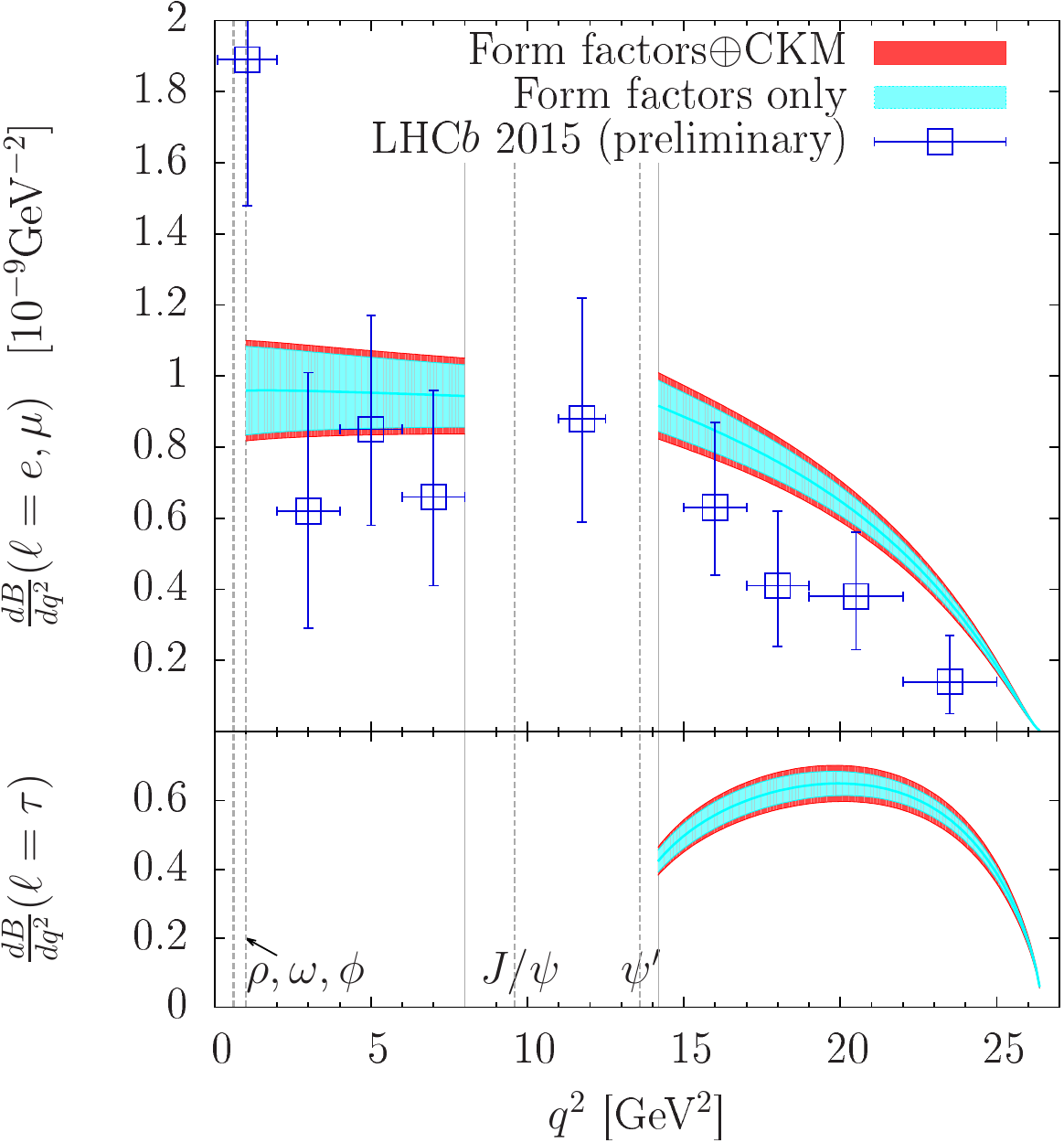}
    \caption{(color online) Partial branching fractions for $B^+\to\pi^+\mu^+\mu^-$ (upper panel) and
        $B^+\to\pi^+\tau^+\tau^-$ (lower panel) outside the resonance regions.
        Different patterns (colors) show the contributions from the main sources of 
        uncertainty; those from the remaining sources are too small to be visible.
        For $B^+\to\pi^+\mu^+\mu^-$, new measurements from LHCb~\cite{TekampeDPF2015}, which were announced after our paper appeared, are overlaid.}
    \label{fig:dBdq2}
\end{figure}

Using $f_T$ from this work and $f_+$ and $f_0$ just described, we show the Standard-Model partial branching
fractions for $B\to\pi\ell^+\ell^-$ in Fig.~\ref{fig:dBdq2}.
Other ingredients are needed besides the form factors.
We take the Wilson coefficients from Ref.~\cite{Huber:2005ig}, the CKM elements from
Ref.~\cite{Charles:2004jd}, the meson masses and lifetimes from Ref.~\cite{Agashe:2014kda}, and the $b$- and
$c$-quark masses from Ref.~\cite{Ali:2013zfa}.
To calculate contributions that cannot be parameterized by the form factors, we employ QCD factorization at
low $q^2$~\cite{Beneke:1999br,Beneke:2000ry,Beneke:2000wa,Asatrian:2001de,Beneke:2001at,Asatryan:2001zw,%
Asatrian:2003vq,Beneke:2004dp,Bobeth:2007dw} and an operator product expansion (OPE) in powers of
$E_\pi/\sqrt{q^2}$ at large $q^2$~\cite{Grinstein:2002cz,Seidel:2004jh,Grinstein:2004vb,Greub:2008cy,%
Bobeth:2010wg,Beylich:2011aq,Bobeth:2011gi,Bobeth:2011nj}.
Full details will be provided in Ref.~\cite{BtoKPheno}.

\begin{table}
		\caption{Standard-Model predictions for $B^+\to\pi^+\ell^+\ell^-$ partial branching fractions.
			Those for $B^0$ decays can be obtained by multiplying by the lifetime ratio
			$(\tau_{B^0}/\tau_{B^+})/2=0.463$.
			Errors shown are from the CKM elements, form factors, variation of the high and low matching scales, and the 
			quadrature sum of all other contributions, respectively.}
		\label{tab:SM_BRs}
		\begin{tabular}{c@{\quad}l@{\quad\quad}l}
			\hline\hline
			$[q^2_\text{min},q^2_\text{max}]$ & \multicolumn{2}{c}{$10^9 \times \BR(B^+\to\pi^+\ell^+\ell^-)$} \\
			$(\text{GeV}^2)$  & \phantom{4.36(}$\ell=e,\mu$ & \phantom{4.36(}$\ell=\tau$ \\
			\hline
			$[0.1,2.0]$       &  1.81(11,24,6,2)   &   \\
			$[2.0,4.0]$       &  1.92(11,22,6,3)   &  \\
			$[4.0,6.0]$       &  1.91(11,20,6,3)    &  \\
			$[6.0,8.0]$       &  1.89(11,18,5,3)   &   \\
			$[15,17]$         &  1.69(10,13,3,5)  	& 1.11(7,8,2,4) \\
			$[17,19]$     	  &  1.52(9,10,2,4)  	& 1.25(8,8,2,3)  \\
			$[19,22]$         &  1.84(11,11,3,5)   	& 1.93(12,10,4,5)  \\
			$[22,25]$         &  1.07(6,6,3,3)  	& 1.59(10,7,4,4) \\
			\hline
			$[1,6]$           & 4.78(29,54,15,6)   &  \\
			$[15,22]$         & 5.05(30,34,7,15)  & 4.29(26,25,7,12)  \\
			\hline
			$[4m_\ell^2,26.4]$      & 20.4(1.2,1.6,0.3,0.5) &  \\
			\hline\hline
	\end{tabular}
	\end{table}

Table~\ref{tab:SM_BRs} presents numerical predictions for selected $q^2$ bins.
The last error in parenthesis contains effects of parametric uncertainties in $\alpha_s$, $m_t$, $m_b$,
$m_c$; of missing power corrections, taking 10\% of contributions not directly proportional to the form
factors; and of violations of quark-hadron duality, estimated to be 2\% at high-$q^2$~\cite{Beylich:2011aq}.
At low $q^2$, the uncertainty predominantly stems from the form factors; at high $q^2$, the CKM elements
$|V_{td}^*V_{tb}|$ and form factors each contribute similar errors.
Figure~\ref{fig:dBdq2} and Table~\ref{tab:SM_BRs} represent the second main result of this Letter.

In the regions $q^2\lesssim1~\text{GeV}^2$ and $6~\text{GeV}^2\lesssim q^2\lesssim 14~\text{GeV}^2$, 
$u\bar u$ and $c\bar c$ resonances dominate the rate.
To estimate the total $\BR$, we simply disregard them and interpolate linearly in $q^2$ between the QCD-factorization result at $q^2 \approx 8.5~\text{GeV}^2$ and the OPE result at $q^2 \approx 13~\text{GeV}^2$.  While this treatment
does not yield the full branching ratio, it does enable a comparison with LHCb's published result,
$\BR(B^+\to\pi^+\mu^+\mu^-)=23(6)\times 10^{-9}$~\cite{LHCb:2012de}, which was obtained from a similar
interpolation over these regions.
Our result $\BR(B^+\to\pi^+\mu^+\mu^-)=20.4(2.1)\times10^{-9}$ agrees with LHCb, and is more precise than
the best previous theoretical estimate~\cite{Ali:2013zfa} because we use $f_T$ directly, which avoids a
large uncertainty from varying the matching scale~$\mu$.


\emph{Outlook} -- %
The largest uncertainty in our determination of the $B\to\pi$ form factors is the combined error from
statistics with chiral-extrapolation and discretization effects included.
We will be able to reduce these with calculations on the MILC Collaboration's recently generated four-flavor
ensembles with physical light-quark masses~\cite{Bazavov:2012xda}.
LHCb's measurement of $\BR(B^+\to\pi^+\mu^+\mu^-)$ will improve, and Belle~II expects to observe the neutral
decay mode $B^0\to\pi^0\ell^+\ell^-$.
If a deviation from the Standard Model is observed, our form factors can be used to compute other
observables such as asymmetries, thereby providing information about new heavy particles, such as their
masses, spin, and couplings.

Note added: after this paper was submitted for publication, the LHCb experiment announced a new measurement for the $B^+\to\pi^+\mu^+\mu^-$ differential decay rate~\cite{TekampeDPF2015}.  The new results are shown in Fig.~\ref{fig:dBdq2}.  The large difference in the lowest $q^2$ bin is due to the presence of light ($\rho, \omega, \phi$) resonances, whose effects are important but cannot be estimated in a model-independent manner.  Given the present experimental and theoretical uncertainties, it is too early to discern possible new-physics contributions to this process.


\emph{Acknowledgements} -- %
We thank Ulrik Egede and Tobias Tekampe from LHCb for useful correspondence.
Computations for this work were carried out with resources provided by the USQCD Collaboration, the Argonne
Leadership Computing Facility, the National Energy Research Scientific Computing Center, and the Los Alamos
National Laboratory, which are funded by the Office of Science of the United States Department of Energy;
and with resources provided by the National Institute for Computational Science, the Pittsburgh
Supercomputer Center, the San Diego Supercomputer Center, and the Texas Advanced Computing Center, which are
funded through the National Science Foundation's Teragrid/XSEDE Program.
This work was supported in part by the U.S.\ Department of Energy under Grants
No.~DE-FG02-91ER40628 (C.B.), %
No.~DE-FC02-06ER41446 (C.D., L.L., S.-W.Q.), %
No.~DE-SC0010120 (S.G.), %
No.~DE-FG02-91ER40661 (S.G.), %
No.~DE-FG02-13ER42001 (D.D., A.X.K.), %
No.~DE-FG02-91ER40664 (Y.M.), %
and No.~DE-FG02-13ER41976 (D.T.); %
by the National Science Foundation under Grants 
No.~PHY-1067881, No.~PHY-10034278 (C.D., L.L., S.-W.Q..), %
No.~PHY-1417805 (J.L., D.D.), %
and No.~PHY-1316748 (R.S.); %
by the URA Visiting Scholars' program (A.X.K., Y.M.); 
by the MINECO (Spain) under Grant FPA2013-47836-C3-1-P, and the Ram\'{o}n y Cajal program (E.G.); %
by the Junta de Andaluc\'{i}a (Spain) under Grants FQM-101 and FQM-6552 (E.G.); %
by the European Commission under Grant No.~PCIG10-GA-2011-303781 (E.G.); %
by the German Excellence Initiative, the European Union Seventh Framework Programme under grant
agreement No.~291763, and the European Union's Marie Curie COFUND program (A.S.K); %
and
by the Basic Science Research Program of the National Research Foundation of Korea (NRF) funded by the 
Ministry of Education (No.~2014027937) and the Creative Research Initiatives Program (No.~2014001852) of 
the NRF grant funded by the Korean government (MEST) (J.A.B).
Brookhaven National Laboratory is supported by the U.~S.~Department of Energy under Contract
No.~DE-SC0012704. 
Fermilab is operated by Fermi Research Alliance, LLC, under Contract No.~DE-AC02-07CH11359 with the U.S.\
Department of Energy.


\bibliographystyle{apsrev4-1} 
\bibliography{B2Pill,asqtad_ensemb}

\end{document}